# Unconventional phase selection in high-driven systems: A complex metastable structure prevails over simple stable phases

Z. Ye,[1] F. Zhang,[1] Y. Sun,[1,3] M. C. Nguyen,[1] M. I. Mendelev,[1] R. T. Ott,[1] E. Park,[1] M. F. Besser,[1] M. J. Kramer,[1] Z. Ding,[3] C.-Z. Wang,[1] K.-M. Ho[1,2,3,*]

[1]Ames Laboratory, US Department of Energy, Ames, Iowa 50011, USA

[2]Department of Physics and Astronomy, Iowa State University, Ames, Iowa 50011, USA

[3]Hefei National Laboratory for Physical Sciences at the Microscale and Department of Physics, University of Science and Technology of China, Hefei, Anhui 230026, China

[*] Email: kmh@ameslab.gov

**Phase selection in deeply undercooled liquids and devitrified glasses during heating involves complex interplay between the barriers to nucleation and the ability for these nuclei to grow. During the devitrification of glassy alloys, complicated metastable structures often precipitate instead of simpler, more stable compounds. Here, we access this unconventional type of phase selections by investigating an Al-10%Sm system, where a complicated cubic structure first precipitates with a large lattice parameter of 1.4 nm. We not only solve the structure of this "big cubic" phase containing ~140 atoms but establish an explicit interconnection between the structural orderings of the amorphous alloy and the cubic phase, which provides a low-barrier nucleation pathway at low temperatures. The surprising rapid growth of the crystal is attributed to its high tolerance to point defects, which minimize the short-scale atomic rearrangements to form the crystal. Our study suggests a new scenario of devitrification, where phase transformation proceeds initially without partitioning through a complex intermediate crystal phase.**

As-quenched amorphous alloys produced by melt spinning or thin-film sputtering display a complicated behavior of devitrification when the samples are gradually heated up, which is controlled by not only the driving force (the free energy difference between the amorphous and crystalline phases), but also atomic diffusivities as well as the topological connection between the amorphous and crystalline phases. For this reason, an amorphous alloy does not necessarily devitrify into the thermodynamically stable phases, although that would have the largest decrease in the free energy. Instead, meta-stable intermediate phases can appear.[1,2,3]

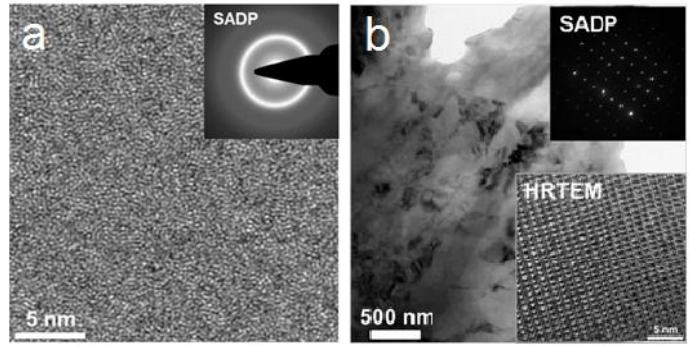

Figure 1. High-resolution transmission electron microscopy (HR TEM) images of amorphous Al-10%Sm ribbons and the 'big cubic' crystal. a, HR TEM image of as-quenched melt-spun ribbons produced at a wheel speed of 30 m/s. The inset shows the selected area diffraction pattern (SADP) that exhibits a featureless microstructure and a diffuse ring typical of an amorphous phase. b, Bright field (BF) showing grains > 200 nm and HR TEM image of the 'big cubic' crystal showing a well ordered lattice after heating to 456K. The inset shows the corresponding SADP.

The Al-Sm binary system can serve as a model system for marginal glass-forming alloys given that it forms an amorphous phase over a wide range of composition.[1,4,5] When as-quenched amorphous Al-10%Sm melt-spun ribbons (Fig. 1a) are isochronally heated, they show a multi-step devitrification pathway that is characterized by a number of meta-stable crystalline phases, as identified by X-ray diffraction (XRD) measurements. The big cubic phase (termed MS1 in Ref. [4,5,6]) is the first phase to appear during devitrification, with multiple crystal grains appearing after a hold of 456 K for 10 min. They continue to grow rapidly until the nearly perfect polymorphic transformation completes in 1 hour, as shown in Fig. 1b. Why such a complex crystal structure containing ~140 atoms is selected over other simpler and more stable crystal structures remained unsolved for 20 years. This mysterious behavior of phase selections is not rare in the crystallization process of glassy alloys. A big-cube-type phase including 96 atoms precipitates in the crystallization of several Zr- and Hf-based glassy alloys such as Zr(Hf)-Al-Ni-Cu systems.[7,8,9] Quasi-crystals, which can be approximated by complex crystal structures with large lattices, are also frequently seen as the primary precipitation phase in a number of glassy alloy systems.[7,10,11] Achieving a fundamental understanding of these unconventional phase selections will provide access to a class of metastable structures which are not reachable by other material processing methods.

A thorough understanding of the surprising phase selections is not possible without the identification of the precipitation phases, which are often complicated due to their metastability, large unit cells, and the presence of point defects such as anti-sites and vacancies. Typically neutron or X-ray powder diffraction, sometimes in concert with transmission electron diffraction, is used to solve crystalline structures. However, random point defects commonly seen in the intermediate metastable phases present a major difficulty for structure identification. In this paper, we apply a new approach that integrates an efficient genetic algorithm (GA) with experimental diffraction data to solve the big cubic crystal structure. A detailed discussion of the approach, including its schematic flow diagram, is given in the Supplementary Information.

The big cubic phase has long been recognized as cubic with $a \sim 9.8$ Å.[6] However, we find that the primitive cubic cell cannot index all the peaks (see Fig. S2 in the Supplementary Information). Instead, a body-centered cubic cell with $a \sim 14$ Å has peaks that perfectly match with experiment in terms of the peak positions. The estimated number of atoms per unit cell is around 140 assuming a density similar to the glass of 0.051 atoms/Å$^3$. Even though we have identified the lattice, the permutations for decorating such a large unit cell are overwhelming. To aid the identification of a realistic atomic decoration, we apply a GA for the global search. Genetic algorithms,[12,13] based on the principles of natural evolution, i.e., survival of the fittest, involve operations that are analogues of the evolutional processes of crossover, mutation and natural selection. By defining fitness as a function of formation energy, we can reliably locate favorable low-energy structures, which had been successfully demonstrated in predicting crystal structures.[14,15,16,17]

We performed a GA search using experimental Sm concentration (10 %) and did not obtain any satisfactory agreement with the experiment. Therefore, we allowed for a search over a wider composition range. In addition, to expedite the GA search, we used a classical interatomic potential in the Finnis-Sinclair (FS) formalism[18,19] for energy calculations. We calculated the XRD pattern of each structure in the resulting pool using RIETAN-FP[20] and a profile factor $F_{XRD}$ is defined, as described below, to assess the overall agreement between the calculated and experimental XRD patterns:

$$F_{XRD} = \sum_n \max(I_{n,cal}, I_{n,exp}) \times \left(\log \frac{I_{n,cal}}{I_{n,exp}}\right)^2 \qquad (1)$$

where $I_{n,cal}$ and $I_{n,exp}$ are the calculated and measured intensities of the $nth$ normalized XRD peak, respectively. The structure with the best $F_{XRD}$ has the configuration of $Al_{120}Sm_{22}$ with space group $I\,m\,\bar{3}\,m$ (No. 229) with 6 unique Wyckoff positions (Table 1).

| Lattice parameters (in unit of Å) | | | | | |
|---|---|---|---|---|---|
| $a = 13.904\ (14.06)$ | | | | | |
| Atomic coordinates and site occupancies | | | | | |
| | x | y | z | Wyckoff | Occupancies |
| Al(1) | 0 | 0.143 (0.142) | 0.316 (0.302) | 48j | 1 |
| Al(2) | 0.849 (0.842) | 0 | 0 | 12e | 1 (0.75) |
| Al(3) | 0.25 | 0 | 0.5 | 12d | 1 |
| Al(4) | 0.171 (0.164) | 0.171 (0.164) | 0.406 (0.405) | 48k | 1 |
| Sm(1) | 0 | 0.5 | 0.5 | 6b | 1 |
| Sm-Al | 0.335 (0.328) | 0.335 (0.328) | 0.335 (0.328) | 16f | 1/0 (0.5/0.5) |

Table 1. Lattice parameters and atomic coordinates of the $Al_{120}Sm_{22}$ phase based on the DFT and Rietveld fitted values to the experimental observed phase in parenthesis. The fitting of the experimental data is wRp = 0.0593 and Rp = 0.0455.

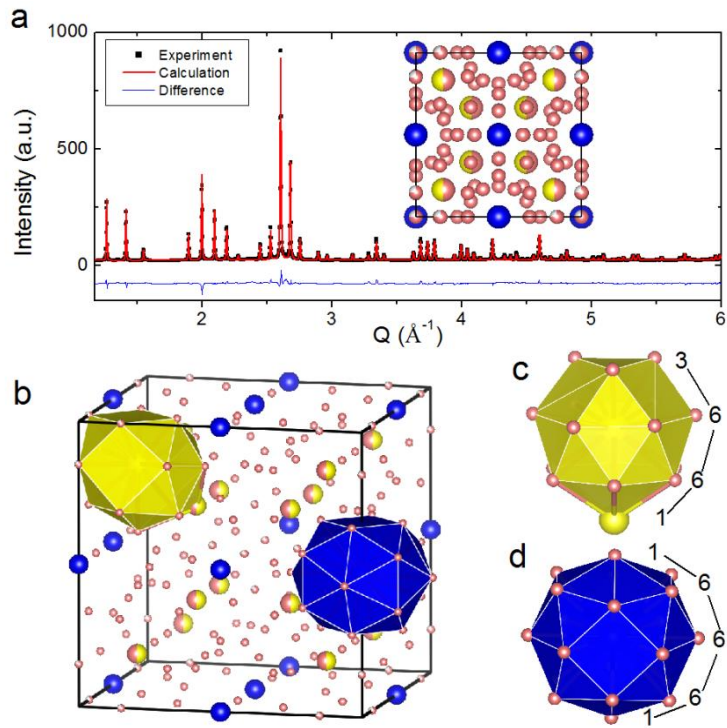

Figure 2. Structure of refined $Al_{125}Sm_{14}$ and its XRD pattern. a, Fitted XRD pattern of $Al_{125}Sm_{14}$. Inset: structure of refined $Al_{125}Sm_{14}$. Blue/yellow and pink represent Sm and Al atoms, respectively. b, Refined $Al_{125}Sm_{14}$ is made up of 2 Sm-centered motifs: c, the 3-6-6-1 motif, and d, the 1-6-6-6-1 motif. The 3-6-6-1 motif is the dominate motif in the undercooled liquid.

Since the crystal structure with the best $F_{XRD}$ has the Sm concentration (15.5%) which is considerably different than the experimental one (10%), a Rietveld refinement[21,22] is performed to determine the site occupancies and anti-site defects. Standard Rietveld analysis was carried out refining lattice parameter, atomic coordinates, site occupancies, thermal parameters, peak shape profiles, etc. The structure after Rietveld refinement is shown in the inset of Fig. 2a, and its XRD pattern compared with experiment is shown in Fig. 2a. The refined atomic coordinates and fitted occupancy of each site are listed in Table 1. Excellent match with experiment is achieved when the Al(2) site (12e) is 75% occupied and the Sm/Al site (16f) is evenly shared by Sm and Al (assuming Sm/Al site to be fully occupied by Sm alone has a much poorer fit). To distinguish the fully occupied (6b) and shared (16f) Sm sites, they are marked in Fig. 2a,b with blue and yellow, respectively. Based on the packing of the atoms around these Sm atoms, the Sm-centered clusters are named as 3-6-6-1 and 1-6-6-6-1 motifs, respectively, as shown in Fig. 2b-d.

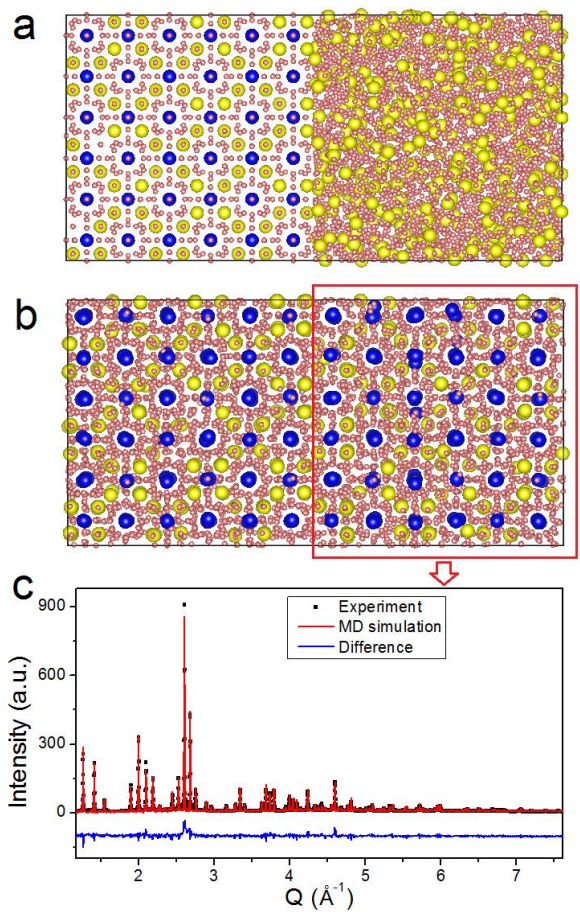

Figure 3. MD simulation results. a, Refined structure $Al_{125}Sm_{14}$ in Fig. 2 serves as a seed (left part) to grow crystalline structure from undercooled Al-10%Sm liquids (right part). b, The crystal structure is fully grown within 300 ns. The Sm atoms (blue) outline the cubic frame, with randomly distributed Sm-Al anti-sites (yellow) embedded in. c, The XRD pattern of the grown structure compared with experiment.

While the crystal structure described above is in excellent agreement with the experimental XRD data, it does not really prove that it is realistic. It is well known that over-fitting (i.e., fit to noise or experimental error by utilizing too many fitting parameters) can occur in Rietveld analysis, which can lead to unphysical solutions of the structures. Moreover, it is not clear why such a complicated structure with the unit cell containing ~140 atoms can be formed from the disordered matrix instead of simpler and more stable phases. To address these questions we perform a large scale molecular dynamic (MD) simulation to study the liquid-to-crystalline transformation using the same classical potential that successfully located the $Al_{120}Sm_{22}$ structure. The initial model contains the liquid $Al_{90}Sm_{10}$ alloy and a seed of the refined $Al_{125}Sm_{14}$ structure with random vacancies and anti-site defects, as predicted by Rietveld analysis (Fig. 3a). Since the MD simulation of the crystal growth can be very computationally expensive, it is performed at an elevated temperature of 800 K to further expedite the transformation (the devitrification of this alloy is experimentally observed at ~ 500 K). The crystal structures are fully grown within 300 ns as shown in Fig. 3b (a short movie is provided in the Supplementary Information). We calculate the XRD pattern of the grown crystal structure containing ~3800 atoms and compare with experimental data. With no need of fitting occupancies or atomic positions, the XRD pattern matches excellently with experiment (Fig. 3c), indicating that the grown structure is realistic.

Figure 4 shows a comparison between the structures obtained from the Rietveld analysis and MD simulation. The simulation shows that the backbone of the structure is the cubic frame of Sm atoms (blue). This is the Sm sites for which the Rietveld analysis predicted the occupancy of 1. All other sites can host point defects (vacancies, interstitials and anti-sites). Analysis of Fig. 4 shows that the refined structure given by Rietveld analysis represents an overall averaging of the spontaneously grown defects in MD simulations and the grown structure (Fig. 4b) captures the detailed characteristics of the defects, which are not clear in the Rietveld refined structure (Fig. 4a).

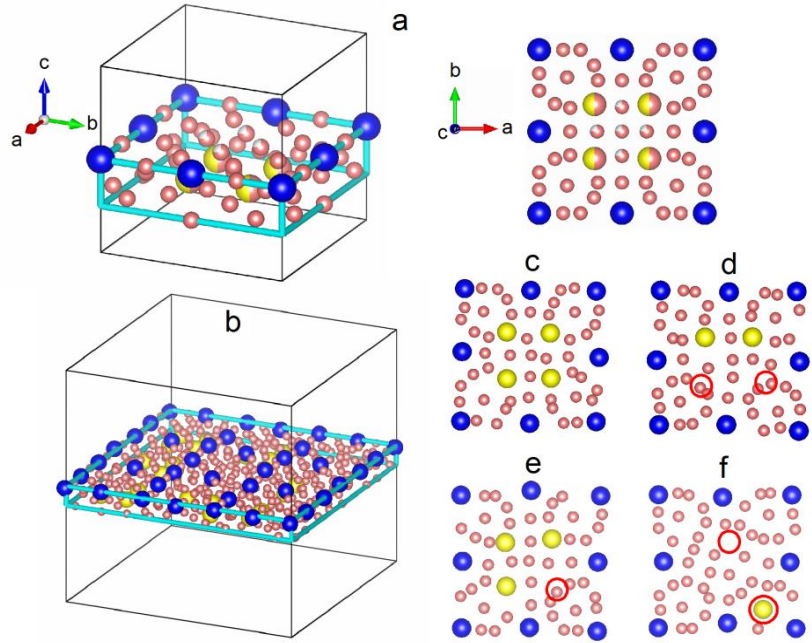

Figure 4. A comparison of defects suggested by Rietveld analysis and by MD simulations. a, A slice with a thickness of $a/4$ of the structure suggested by Rietveld analysis. b, A slice with the same thickness from the grown $3 \times 3 \times 3$ supercell structure. c-f, A close look of the vacancies and anti-site defects as marked in red circle.

We now discuss why the polymorphic phase transformation from the amorphous state to a metastable cubic phase described above is preferred over a transformation to simpler and more stable phases (e.g., fcc Al and $Al_{11}Sm_3$). In our MD simulations, the crystal structures are fully grown within 300 ns. This high growth speed for a structure with such a large unit cell is remarkable. In contrast, the thermodynamically stable and structurally simpler phase, fcc-Al, hardly grows under similar simulation conditions. One obvious advantage of the polymorphic transition is that it does not require partitioning. Therefore, this transformation can be completed before simpler and more stable phases form, which requires compositional segregation with long-range atomic rearrangements. However, the MD simulation shows that even very simple B2 phase (containing only 2 atoms in the unit cell) grows from the liquid several orders of magnitude slower than the rate of solidification in pure metals.[23] The B2 phase represents a typical binary system in which anti-site defects are thermodynamically unfavorable. Whenever an anti-site defect occurs at the solid liquid interface (SLI), the local atomic rearrangements needed to correct the defect are very time-consuming. In contrast, we have demonstrated that the big cubic phase is tolerant to a large amount of Al/Sm anti-site defects, which in turn allows a high SLI mobility. TEM analysis of quenched glass heated to a partially devitrified state shows low number of ~ 200 nm crystallites supporting low nucleation and fast growth which supports this hypothesis.[5]

Another factor that contributes to the surprising fast growth of the big cubic phase out of a liquid matrix is the clear interconnection between the nanoscale structural orderings in the undercooled liquid and the described crystal phase here. Mostly it is seen in the Sm centered clusters. 57% of the clusters in the big cubic phase are made of the 3-6-6-1 motif (Fig. 2b,c). From our recent work,[25] the same motif is the most frequently seen in the undercooled Al-10%Sm liquids. In that work, the most common energetically favorable packing motifs in crystalline structures serve as templates and are compared to the atomic clusters in the actual liquids using a cluster-alignment method. Among them the 3-6-6-1 motif was found to be the dominant motif in the undercooled liquids. This interconnection indicates a low-barrier pathway for the formation of this crystal phase from the amorphous state. Analogous interconnections are also seen in other glassy systems. For example, the precipitated nano-icosahedral phases have long been believed to correlate with the possible local icosahedral order in Zr- and Hf-based glassy alloys.[7] Here, we establish an explicit correlation between the glassy and crystal states sharing a well-described sub-unit cell motif (Fig. 2c).

In summary, using the high resolution X-ray diffraction data we are able to describe the previously unidentified metastable crystal phase, which appears during a polymorphic phase transformation from the amorphous Al-10%Sm alloy. Rietveld analysis demonstrates that the experimental data can be explained only with the assumption of a considerable amount of point defects contained in the crystal phase. Moreover, we perform MD simulations and grow this crystal phase from a seed in the liquid alloy. Our study suggests a new scenario of devitrification, where phase transformation proceeds initially without partitioning through a complex intermediate crystal phase. In our present scenario, the similarity between the crystal and undercooled liquid structures favors a few quenched-in nuclei in the amorphous state that can grow once the temperature is sufficiently high enough. The surprisingly fast growth of the crystal phase with such a large unit cell can be partially attributed to the tolerance of the crystal phase to point defects and the structural similarity between the crystal and glass, which minimizes the short-range atomic rearrangements to grow the crystal. Achieving a fundamental understanding of the crystallization process in glassy alloys, which is controlled by not only the driving force, but also atomic diffusivities as well as the interconnection between the structural orderings in the amorphous and crystalline phases, will allow us to develop a whole new class of materials through control and manipulation of the devitrification process.

**Methods**

**Experimental procedure.** Ingots of Al90Sm10 were prepared by electric arc melting under Ar atmosphere from highly pure Al (99.99 wt.%) and Sm (99.9 wt.%) elements. Amorphous ribbons were produced from bulk alloy by a Cu chill block single-roller melt-spinning technique under Ar atmosphere at tangential speeds of 30 m/s. High temperature X-ray diffraction studies were carried out using high-energy transmission synchrotron X-ray diffraction (HEXRD). More details can be found in the Supplementary Information.

**Genetic algorithm.** A GA is used to search for low energy structure by defining the fitness as a function of energy. All structure relaxations during the GA search are performed by LAMMPS code[26] with Embedded-Atom Method (EAM) potential in Finnis-Sinclair form.[18] As previously shown,[16,19] this FS potential fitted to first principles calculation data, in general, gives a satisfactory estimation of the relative thermodynamic stability of the known stable and meta-stable phases.

**Density functional theory (DFT).** All DFT[27] calculations are performed using the Vienna *ab initio* simulation package[28] (VASP) with projector-augmented wave (PAW) pseudopotential method[29,30] within generalized-gradient approximation (GGA)[31].

**Fitting XRD pattern.** A standard Rietveld analysis was carried out using the GSAS package and the EXPGUI interface[32] to refine the $Al_{120}Sm_{22}$ structure from GA search. Lattice parameter, atomic coordinates, site occupancies, thermal parameters, and peak shape profiles are refined to get the XRD pattern in Fig. 2a. To get the XRD pattern of the grown crystal structure from MD simulation in Fig. 3c, we first use Diamond software to get the XRD peak intensities from a snapshot of the grown crystal containing ~3800 atoms and manipulate them with a Gaussian function for each reflection. The lattice parameter and the Gaussian RMS width are optimized with respect to experimental data.

**Molecular dynamic simulations of the crystal-liquid interface.** The classical MD simulations are performed with the same potential in the Finnis-Sinclair form.[18] The constant number of atoms, pressure and temperature (*NPT*) ensemble is applied with a time step of 3 *fs* to update the system configuration. The liquid sample is created with 3458 Al atoms and 381 Sm atoms. Randomly generated configurations with cubic supercell are equilibrated at 2000 K over 1 ns. The sample is cooled down to 800 K with a cooling rate of $1\times10^{10}$ K/s. Then the liquid sample is put in contact with the crystalline seed to anneal at 800K for 285 ns to finish the liquid-crystal interface simulations.

**Acknowledgements**

We would like to thank Jon Almer for his assistance in the HEXRD experiments. Work at Ames Laboratory was supported by the US Department of Energy, Basic Energy Sciences, Division of Materials Science and Engineering, under Contract No. DE-AC02-07CH11358, including a grant of computer time at the National Energy Research Supercomputing Center (NERSC) in Berkeley, CA. The high-energy X-ray experiments were performed at the XOR beamline (sector 1) of the Advanced Photon Source, Argonne National Laboratory, under Grant No. DE-AC02-06CH11357.

**Contributions**

K.M.H. and C.Z.W. designed and supervised the project. Z.Y., F.Z., Y.S. and M.C.N. performed GA search. M.I.M developed the Finnis-Sinclair potential for the Al-Sm system. M.J.K. performed Rietveld refinement. Z.Y., Y.S. and Z.D. performed MD simulation. R.T.O., E.P., M.F.B. and M.J.K. performed experimental measurements of the XRD pattern. Z.Y., F.Z., M.I.M., C.Z.W. and K.M.H. coordinated the work on the manuscript with contributions from other authors.

**Competing financial interests**

The authors declare no competing financial interests.

# Supplementary Information

# Unconventional phase selection in high-driven systems: A complex metastable structure prevails over simple stable phases

Z. Ye,[1] F. Zhang,[1] Y. Sun,[1,3] M. C. Nguyen,[1] M. I. Mendelev,[1] R. T. Ott,[1] E. Park,[1] M. F. Besser,[1] M. J. Kramer,[1] Z. Ding,[3] C. –Z. Wang,[1] K. –M. Ho[1,2,3,*]

[1]Ames Laboratory, US Department of Energy, Ames, Iowa 50011, USA

[2]Department of Physics and Astronomy, Iowa State University, Ames, Iowa 50011, USA

[3]Hefei National Laboratory for Physical Sciences at the Microscale and Department of Physics, University of Science and Technology of China, Hefei, Anhui 230026, China

**Methodological Discussion**

We schematically depict our integrated approach in Figure S1. We first use the experimental diffraction patterns to deduce the lattice parameters and the space group symmetry. Then, we perform a global search for low energy structures using a genetic algorithm (GA), with the constraints of the pre-determined space group symmetry, which effectively reduces the size of the system and at the same time restrict the GA search to a subset of metastable structures that satisfy the experimental data. Even with the help of the symmetrization, it remains impractical to use straightforward first-principles calculations throughout the search for large unit cells. Thus, we use an efficient classical potential for energy calculations during the GA search. To ensure its accuracy, the classical potential has been pre-adapted in GA searches on smaller cells, using procedures described in Ref. 14. The XRD patterns of the structures in the converged GA pool are then calculated and compared with experiments. Those structures with apparently inconsistent XRD patterns with experiments are screened out of the pool. Since the full space group symmetry is enforced throughout the search, the final structures have full occupancies for all the Wyckoff positions. In order to account for possible random vacancies or anti-sites associated with partial occupancies, the best candidates from the XRD screening are fed to a Rietveld analysis, in which the lattice parameters, atomic coordinates, and the site occupancies are fine-tuned to obtain an optimal match between the experimental and calculated diffraction patterns. Since physical guidelines are limited in the Rietveld analysis, to ensure that it does not result in unphysical results, we finally perform a large scale molecular dynamics simulation on the transformation from the amorphous to the crystalline phases. The newly grown crystalline phase can help confirm that whether the vacancies or anti-sites predicted in the Rietveld analysis are realistic.

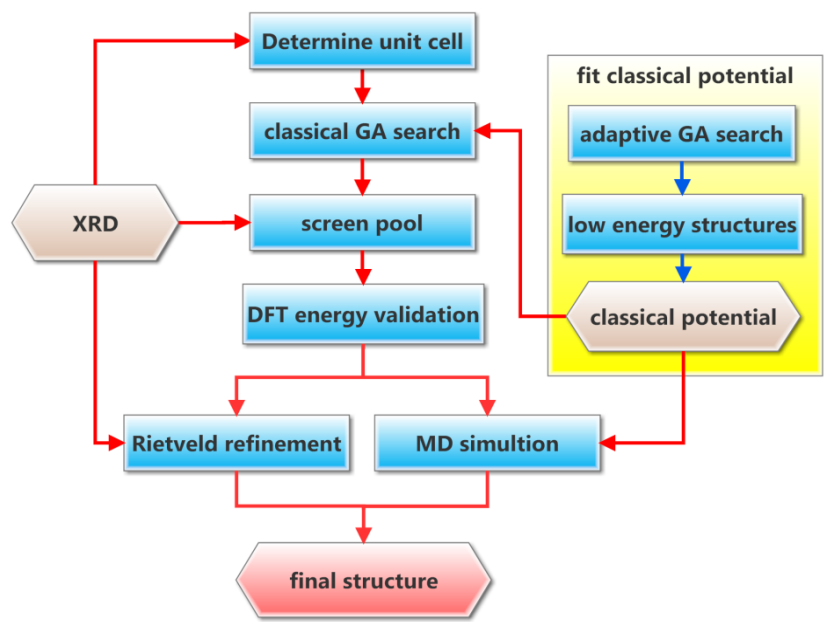

Figure S1. A schematic flow diagram of our integrated approach.

**Determine the unit cell**

The big cubic phase has long been recognized as cubic with $a \sim 9.8$ Å.[6] We look to move beyond this limited description by revisiting the phase selection in Al-Sm alloys produced by melt spinning. By using higher resolution in-situ time-resolved XRD at the Advanced Photon Source (APS), we find that the primitive cubic cell cannot index all the peaks (Fig. S2). The missing peaks are all minor and can be easily overlooked unless XRD patterns with very high resolution are acquired. Using standard space group peak matching techniques, we find that a body-centered cubic cell with $a \sim 14$ Å have XRD peaks that perfectly match with experiment in terms of the peak positions.

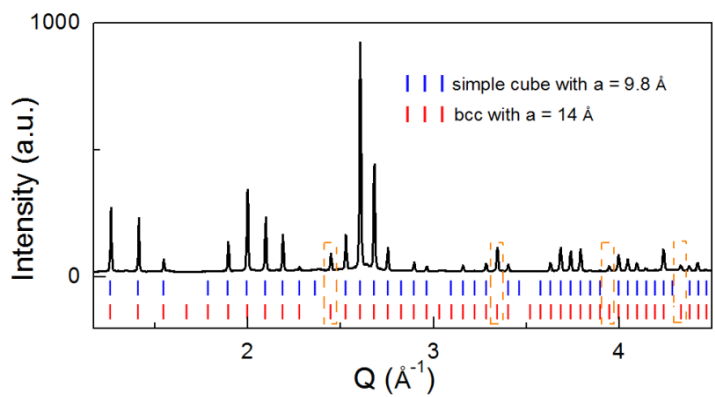

Figure S2. Experiment XRD of the big cubic phase. The tickmarks are for primative cubic cell with $a = 9.8$ Å (blue) and bcc with $a = 14$ Å (red). The orange frames indicate the missing minor peaks for simple cube with $a = 9.8$ Å.

**Experimental procedure**

Ingots of $Al_{90}$-$Sm_{10}$ (at.%) were prepared by arc melting under an He atmosphere from highly pure Al (99.99 wt.%.) and Sm (99.9 wt.%.) elements. Amorphous ribbons with a thickness of 20-30 μm were fabricated by a Cu chill block single-roller melt-spinning technique under He atmosphere at a wheel speed of 30 m/s.

The thermal properties of as-prepared samples were determined by a differential scanning calorimetry (DSC, Perkin Elmer Pyris 1) at a constant heating rate of 10 K/min. The as-prepared and devitrification product phases were investigated using *in-situ* high-energy synchrotron wide-angle X-ray scattering (WAXS) at the Advanced Photon Source at Argonne National Laboratory in collaboration with the Midwest Universities Collaborative Access Team beam line 6ID-D. The as-prepared samples were cut and inserted into 2 mm diameter thin-walled quartz tubes that were evacuated and sealed in He to avoid oxidation during heating. The sealed capillary tubes were exposed to 80 keV X-rays of wavelength 0.01536 nm during *in situ* devitrification experiments. The diffraction data were collected with Debye-Scherrer geometry by a CCD camera every 6 s.

The as spun alloy and annealed at 456 K for 1 hr were examined by a scanning/transmission electron microscope (S/TEM, FEI Tecnai $G^2$ F20). An ion beam milling method was used to prepare the thin foil specimens of as-prepared and annealed samples. Specimens for High resolution TEM (HRTEM) analyses were thinned using ion milling at 150 K with a low energy of 1.0 kV.